\documentstyle[12pt,aasms4]{article}

\begin{document}

\title{X-ray Nova XTE J1550-564: RXTE Spectral Observations}

\author{Gregory J. Sobczak\altaffilmark{1}, Jeffrey E. McClintock\altaffilmark{2},
Ronald A. Remillard\altaffilmark{3}, Alan M. Levine\altaffilmark{3}, Edward H. 
Morgan\altaffilmark{3}, Charles D. Bailyn\altaffilmark{4}, and Jerome A.
Orosz\altaffilmark{5}}

\altaffiltext{1}{Harvard University, Astronomy Dept., 60 Garden St. MS-10, Cambridge,
MA 02138; gsobczak@cfa.harvard.edu}
\altaffiltext{2}{Harvard-Smithsonian Center for Astrophysics, 60 Garden St.,
Cambridge, MA 02138; jem@cfa.harvard.edu}
\altaffiltext{3}{Center for Space Research, MIT, Cambridge, MA 02139; rr@space.mit.edu,
aml@space.mit.edu, ehm@space.mit.edu}
\altaffiltext{4}{Dept. of Astronomy, Yale University, P. O. Box 208101, New Haven, CT
06520; bailyn@astro.yale.edu}
\altaffiltext{5}{Dept. of Astronomy and Astrophysics, The Pennsylvania State
University, 525 Davey Laboratory, University Park, PA 16802; orosz@astro.psu.edu}

\begin{abstract} 

Excellent coverage of the 1998 outburst of the X-ray Nova XTE~J1550--564 was provided
by the {\it Rossi X-ray Timing Explorer}.  XTE~J1550--564 exhibited an intense
(6.8~Crab) flare on 1998 September~19 (UT), making it the brightest new X-ray
source observed with RXTE.  We present a spectral analysis utilizing 60 Proportional
Counter Array spectra from 2.5--20~keV spanning 71 days, and a nearly continuous All
Sky Monitor light curve.  The spectra were fit to a model including multicolor
blackbody disk and power-law components.  XTE~J1550--564 is observed in the very high,
high/soft, and intermediate canonical outburst states of Black Hole X-ray Novae.  

\end{abstract}


\section{Introduction}

The X-ray nova and black hole candidate XTE~J1550--564 was discovered (Smith et
al.~1998) with the All Sky Monitor (ASM; Levine et al.~1996) aboard the {\it Rossi
X-ray Timing Explorer} (RXTE) just after the outburst began on 1998 September~6. The
position quickly led to the identification of the counterparts in the optical (Orosz,
Bailyn, \& Jain 1998) and radio (Campbell-Wilson et al.~1998) bands. Early
observations of the source with BATSE (Wilson et al.~1998) revealed an X-ray photon
index in the range of 2.1--2.7; on one occasion the source was detected at energies
above 200~keV.  

The X-ray light curve of XTE~J1550--564 from the ASM is shown in Figure~1. The source
is the brightest transient yet observed with RXTE: the flare of 1998 September~19-20
reached 6.8~Crab (or 1.6~$\times 10^{-7}$~erg~s$^{-1}$~cm$^{-2}$) at 2--10~keV. The
overall profile of the outburst, with its slow 10-day rise, dominant X-ray flare,
30-day intensity plateau, and relatively rapid $\sim$10-day decay timescale (after
MJD~51110), is different from the outbursts of classical X-ray novae like A0620--00
(see Chen, Shrader, \& Livio 1997). The apparent optical and X-ray brightness of
XTE~J1550--564, and even its X-ray light curve, are roughly similar to that of X-ray
Nova Oph 1977, a black hole binary, for which there is a distance estimate of
$\sim$~6~kpc (Watson, Ricketts, \& Griffiths 1978; Remillard et al.~1996).

The discovery of XTE~J1550--564 prompted a series of pointed RXTE observations with
the Proportional Counter Array (PCA; Jahoda et al.~1996) and the High-Energy X-ray
Timing Experiment (HEXTE; Rothschild et al.~1998) instruments.  These were scheduled
almost daily for the first 50 days of the outburst and roughly every two days during
the following two months. The first 14 RXTE observations were part of a guest observer
program with results reported by Cui et al.~(1998). They found that during the initial
X-ray rise (0.7--2.4~Crab at 2--10~keV), the source exhibited very strong QPOs in the
range 0.08--8~Hz. The rapid variability and the characteristics of the X-ray spectrum
suggested that XTE~J1550--564 is powered by an episode of accretion in a black hole
binary system (see Tanaka \& Lewin 1995). The possible presence of a $B\sim22$~mag
counterpart (Jain et al.~1999) is especially important since this may allow radial
velocity studies in quiescence that could confirm the black hole nature of the
primary.  

Herein we present 60 X-ray spectral observations spanning 71 days of the 1998 outburst
of XTE~J1550--564.  These observations include all of our RXTE guest observer program
(\#30191) and the first five public observations of this source.  A timing study based
on these same RXTE observations and observations of the optical counterpart are
presented in companion papers (Remillard et al.~1999 and Jain et al.~1999; hereafter
Paper II and Paper III, respectively).

\section{Spectral Observations and Analysis}

We present 60 observations of XTE~J1550--564 (see Fig.~1) obtained using the PCA
instrument onboard RXTE.  The PCA consists of five xenon-filled detector units (PCUs)
with a total effective area of $\sim$~6200~cm$^{-2}$ at 5~keV.  The PCA is sensitive
in the range 2--60~keV, the energy resolution is $\sim$17\% at 5~keV, and the time
resolution capability is 1~$\mu$sec.  The HEXTE data are not presented here due to
uncertainty in the PCA/HEXTE cross calibration.  

The PCA data were taken in the ``Standard 2'' format and the response matrix for each
PCU was obtained from the 1998~January distribution of response files.  The pulse
height spectrum from each PCU was fit over the energy range 2.5--20~keV, using a
systematic error in the countrates of 1\%, and background subtracted using the standard
background models.  Only PCUs~0~\&~1 were used for the spectral fitting reported here
and both PCUs were fit simultaneously using XSPEC.  

The PCA spectral data were fit to the widely used model consisting of a multicolor
blackbody accretion disk plus power-law (Tanaka \& Lewin 1995; Mitsuda et al.~1984;
Makishima et al.~1986).  The fits were significantly improved by including a smeared
Fe aborption edge near 8~keV (Ebisawa et al.~1994; Inoue 1991) and an Fe emission line
with a central energy around 6.5 keV and a fixed width of 1.2 keV (FWHM).  The fitted
equivalent width of the Fe emission line was $\lesssim100$~eV.  Interstellar
absorption was modeled using the Wisconsin cross-sections (Morrison \& McCammon 1983).
The fitted hydrogen column density varied from 1.7 to 2.2 $\times 10^{22}$~cm$^{-2}$
and was fixed at $2.0 \times 10^{22}$~cm$^{-2}$ in the analysis presented here,
resulting in a total of eight free parameters.  

Three representative spectra are shown in Figures~2a--c.  These spectra illustrate the
range of X-ray spectra for XTE~J1550--564, in which an intense power-law component
dominates a hot disk (a), a strong power-law dominates a warm disk (b), or the disk
dominates a weak power-law (c).  The addition of the Fe emission \& absorption
components is motivated in Figures~3a~\&~b, which show the ratio of a typical spectrum
to the model without and with the Fe emission \& absorption.  The addition of the Fe
emission \& absorption components reduces the $\chi^2_{\nu}$ from 7.9 to 0.9 in this
example.  

The fitted temperature and radius of the inner accretion disk presented here ($T_{in}$
\& $R_{in}$) are actually the color temperature and radius of the inner disk, which
are affected by spectral hardening due to Comptonization of the emergent spectrum
(Shakura \& Sunyaev 1973).  The corrections for constant spectral hardening are
discussed in Shimura \& Takahara (1995), Ebisawa et al.~(1994), and Sobczak et
al.~(1999), but are not applied to the spectral parameters presented here.  The
physical interpretation of these parameters remains highly uncertain.  The model
parameters and component fluxes (see Table~1) are plotted in Figures~4a--e.  All
uncertainties are given at the $1 \sigma$ level.

\section{Discussion of Spectral Results}

The spectra from MJD~51074 to 51113 are dominated by the power-law component which has
photon index $\Gamma =$~2.35--2.86 (Fig.~4c, Table~1).  The source also displays
strong 3--13~Hz QPOs during this time (Paper II), whenever the power-law contributes
more than 60\% of the observed X-ray flux.  This behavior is consistent with the {\it
very high state} of Black Hole X-ray Novae (BHXN) (See Tanaka \& Lewin (1995) and
references therein for further details on the spectral states of BHXN).  After
MJD~51115, the power-law component weakens rapidly, with $\Gamma\sim$~2.0--2.4, and
the disk component begins to dominate the spectrum (see Fig.~2c).  The source
generally shows little temporal variability during this time (Paper II).  We identify
this period with the {\it high/soft state}.  However, during this time the source
occasionally exhibits QPOs at $\sim$~10~Hz (Table~1) and we identify those
observations with the {\it intermediate state}.  The {\it low state} was not observed
and the intensity increased again after MJD~51150.  

From Figure~4b, it appears that the inner radius of the disk does not remain fixed at
the last stable orbit throughout the outburst cycle.  From Table~1, we see that the
intense flare on MJD~51075 is accompanied by a dramatic decrease in the inner disk
radius from 33 to 2~km (for zero inclination ($\theta=0$) and $D=6$~kpc) over one day.
Similar behavior was observed for GRO~J1655--40 during its 1996-97 outburst: the
observed inner disk radius decreased by almost a factor of four during periods of
increased power-law emission in the very high state and was generally larger in the
high/soft state (Sobczak et al.~1999).

The physical radius of the inner disk may vary in these systems, by as much as a
factor of 16 in the case of XTE~J1550--564.  Another possibility, however, may be that
the apparent decrease of the inner disk radius observed during intense flares from
these two sources is caused by the failure of the multicolor disk model at these
times.  This failure could occur when spectral hardening becomes significant in the
inner disk, causing the color temperature to assume a steep radial profile (Shimura \&
Takahara 1995).  In such a case, fitting the multicolor disk model (which assumes $T
\sim r^{-3/4}$) to the resulting spectrum yields an inner disk radius which is smaller
than the physical value (Sobczak et al.~1999).  Thus the actual physical radius of the
inner disk may remain fairly constant in the presence of these intense flares.  

The peak luminosity (bolometric disk luminosity plus 2--100~keV power-law luminosity)
observed during the flare on MJD~51075 is $L=1.2\times10^{39}(D/6kpc)^2$ erg~s$^{-1}$,
which corresponds to the Eddington luminosity for $M=9.6M_{\odot}$ at 6~kpc.

\section{Conclusion}

We have analyzed RXTE data obtained for the X-ray Nova XTE~J1550--564.  Satisfactory
fits to all the PCA data were obtained with a model consisting of a multicolor disk, a
power-law, and Fe emission and absorption components.  XTE~J1550--564 is observed in
the very high, high/soft, and intermediate canonical outburst states of BHXN.  The
source exhibited an intense (6.8~Crab) flare on MJD~51075, during which the inner
disk radius appears to have decreased dramatically from 33 to 2~km (for zero
inclination and $D=6$~kpc).  However, the apparent decrease of the inner disk radius
observed during periods of increased power-law emission may be caused by the failure
of the multicolor disk and the actual physical radius of the inner disk may remain
fairly constant.

\acknowledgements This work was supported, in part, by NASA grant NAG5-3680.  Partial
support for J.M. and G.S. was provided by the Smithsonian Institution Scholarly
Studies Program.  C.B. acknowledges support from an NSF National Young Investigator
award.

\newpage
\begin{figure}
\epsscale{1}
\figurenum{}
\plotfiddle{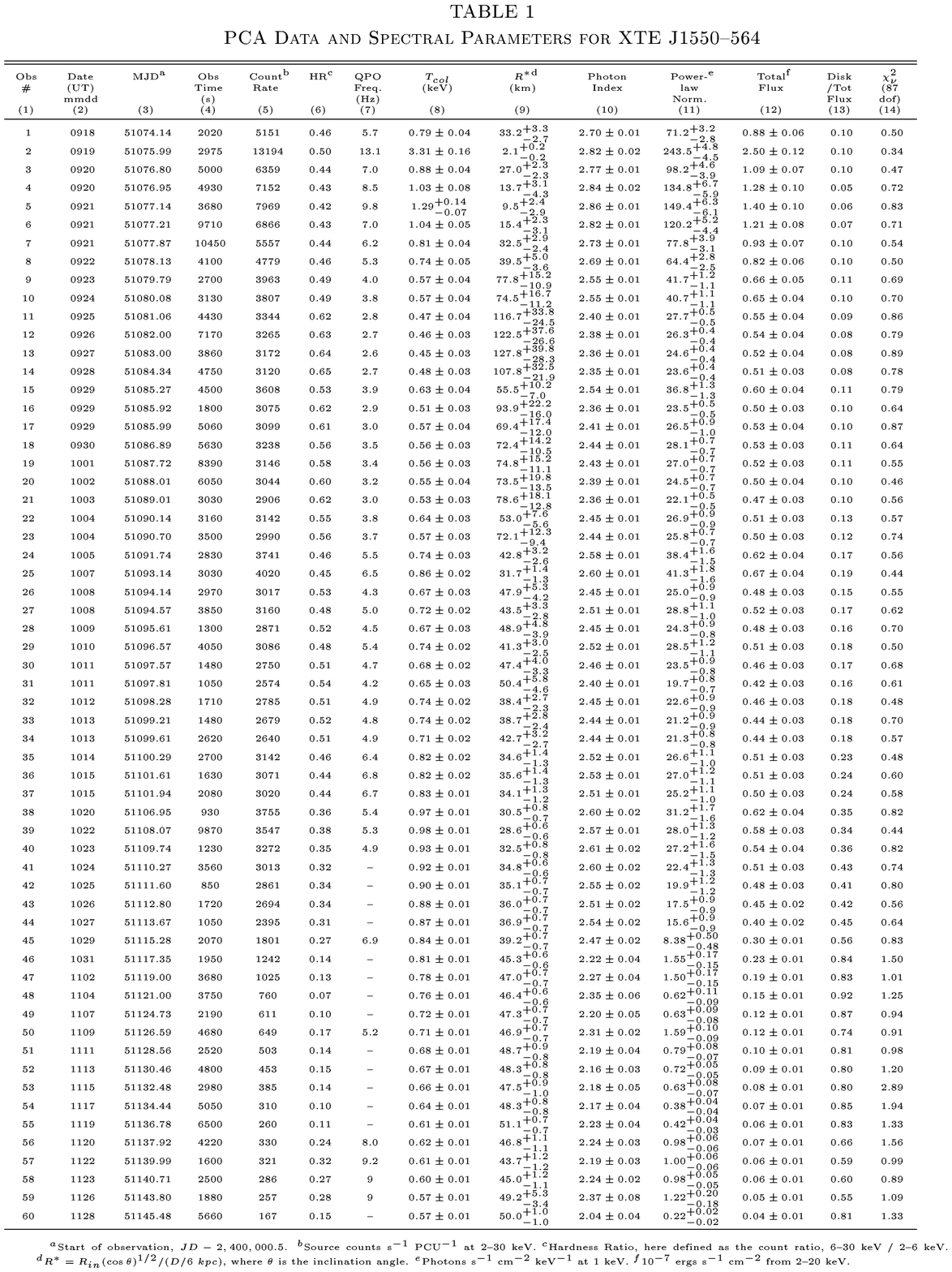}{10pt}{0}{100}{100}{-288}{-375}
\end{figure}

\newpage
\begin{figure}
\figurenum{1}
\epsscale{1}
\plotone{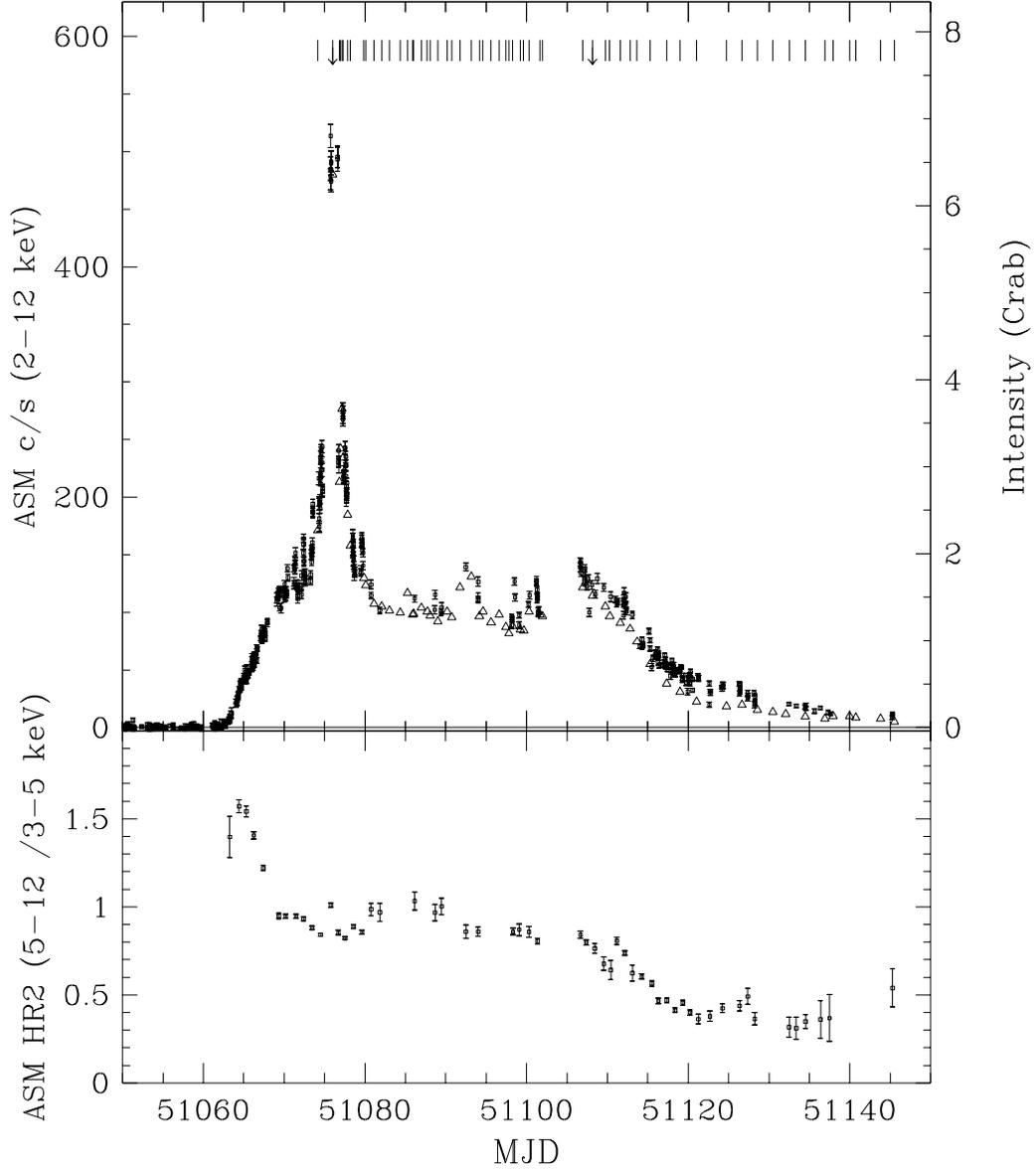}
\caption{(Upper Panel) The 2--12~keV ASM lightcurve and (Lower
Panel) the ratio of the ASM countrates (5--12~keV)/(3--5~keV) for XTE~J1550-564. 
Additional data from the PCA observations are represented by open triangles.  The
small, solid vertical lines in the top panel indicate the times of pointed RXTE
observations; the downward arrows indicate the observations during which the 185~Hz
QPO is present.  The intensity increased again after MJD~51150.  }
\end{figure}

\begin{figure}
\figurenum{2}
\epsscale{0.75}
\plotone{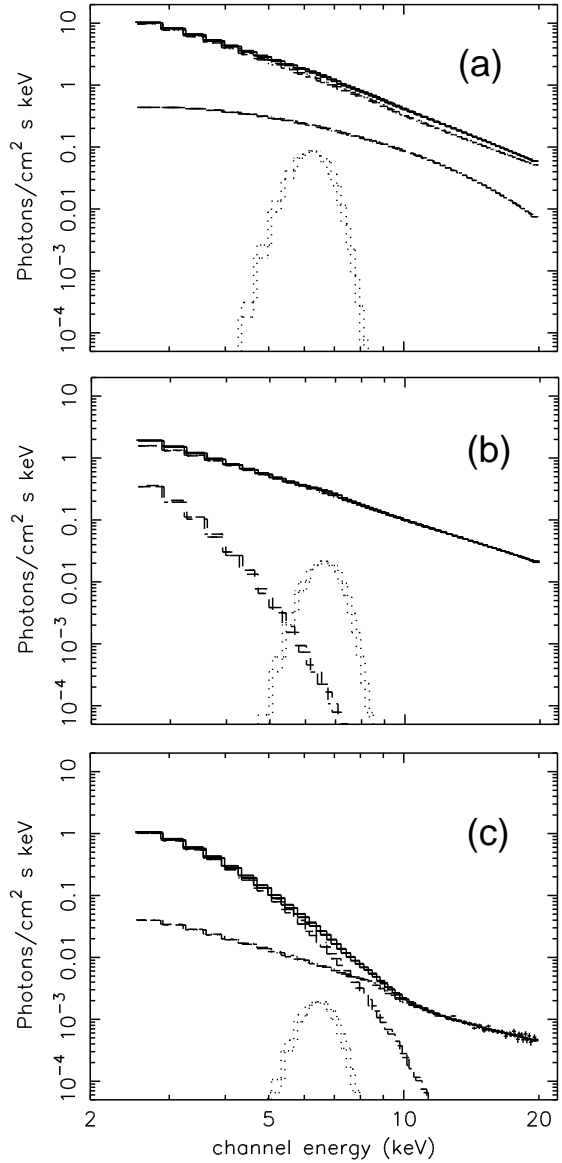}
\caption{Sample PCA spectra from (a) the flare on MJD~51075 (1998
Sept.~19), (b) the very high state on MJD~51083 (1998 Sept.~27), and (c) the high/soft
state on MJD~51121 (1998 Nov.~4).  The individual components of the model are also
shown.  Although error bars are plotted for all the data, they are only large enough
to be visible at the highest energies in panel (c).  }
\end{figure}

\begin{figure}
\figurenum{3}
\epsscale{1}
\plotone{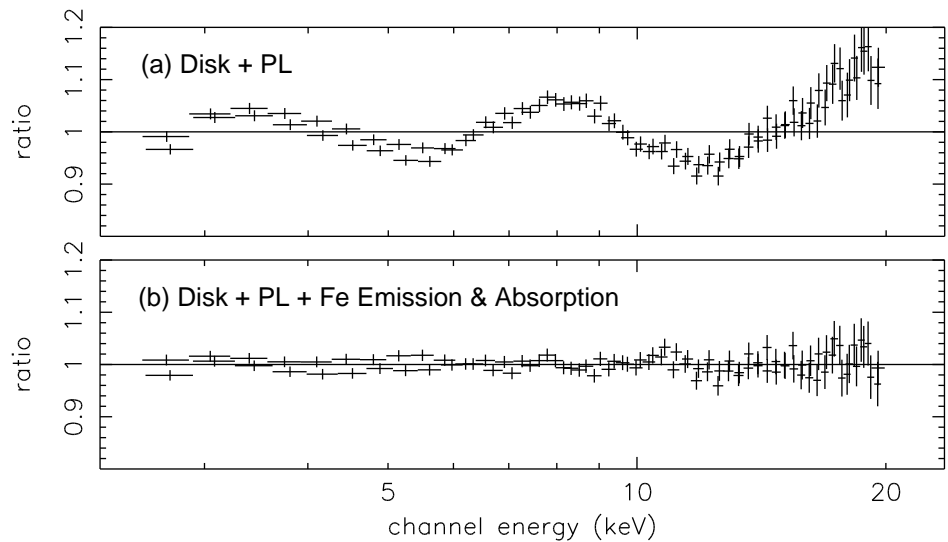}
\caption{The ratio data/model for (a) the multicolor disk plus power-law
model and (b) the multicolor disk plus power-law plus Fe emission \& absorption model
for a representative high/soft state spectrum (MJD~51126, 1998 Nov.~9).  The addition
of the Fe emission \& absorption components improves the $\chi^2_{\nu}$ from (a) 7.9
to (b) 0.9 in this example.  }
\end{figure}

\begin{figure}
\figurenum{4}
\epsscale{1}
\plotone{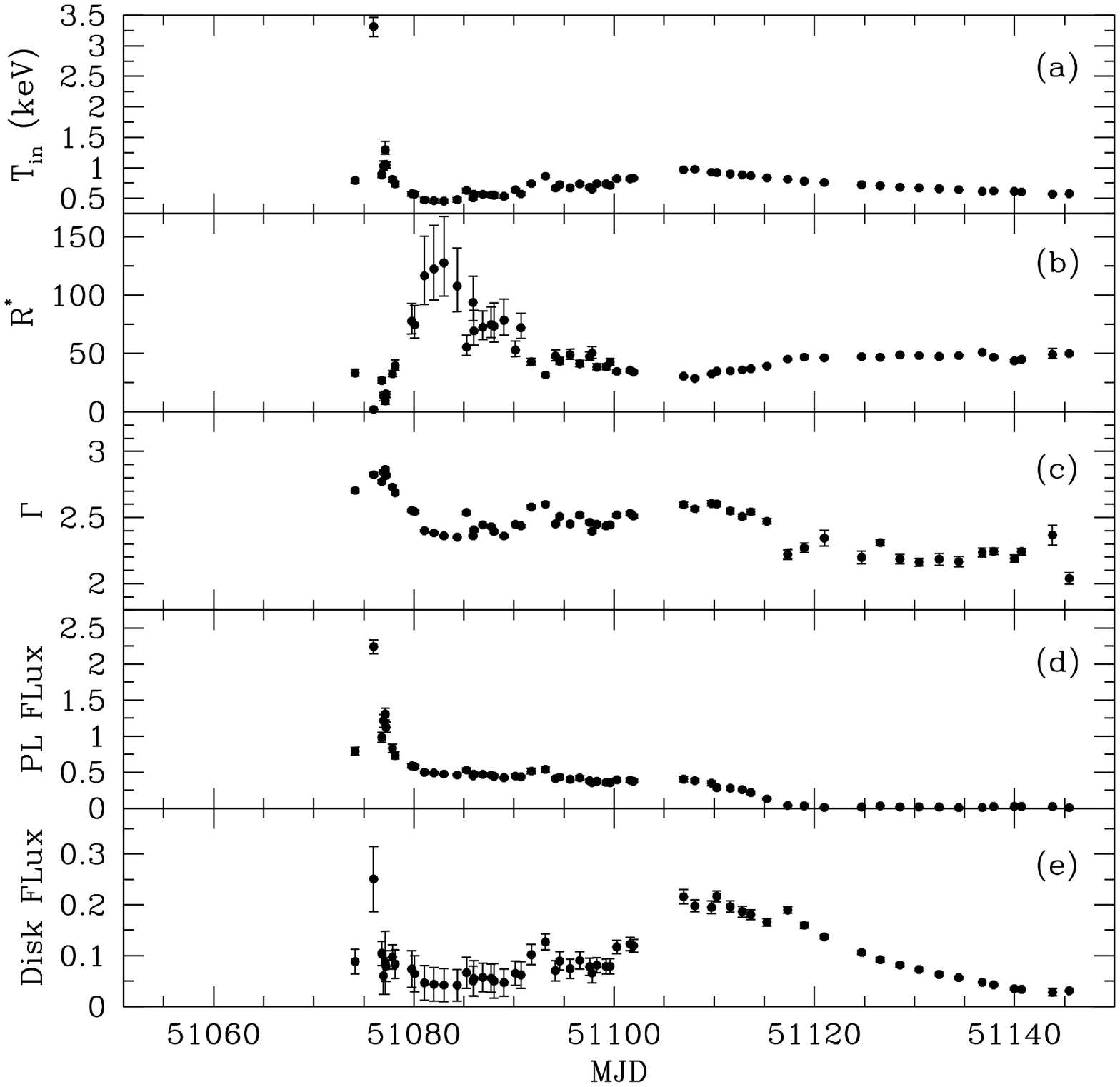}
\caption{Spectral parameters and fluxes for PCA observations of
XTE~J1550--564.  See the text for details on the spectral models and fitting.  The
quantities plotted here are (a) the color temperature of the accretion disk ($T_{in}$)
in keV, (b) the inner disk radius $R^*=R_{in}(\cos \theta)^{1/2}/(D/6~kpc)$, where
$\theta$ is the inclination angle, (c) the power-law photon index $\Gamma$, (d) the
unabsorbed 2--20~keV power-law flux in units of $10^{-7}$~erg~s$^{-1}$~cm$^{-2}$, and
(e) the unabsorbed 2--20~keV disk flux in the same units.  When error bars are not
visible, it is because they are comparable to or smaller than the plotting symbol.  }
\end{figure}


\newpage
\begin{references}
\reference{}Campbell-Wilson, D., McIntyre, V., Hunstead, R., \& Green, A. 1998, \iaucirc
7010
\reference{}Chen, W., Shrader, C. R., \& Livio, M. 1997, \apj, 491 312
\reference{}Cui, W., Zhang, S. N., Chen, W., \& Morgan, E. H. 1999, \apj, 512, L43
\reference{}Ebisawa, K., Ogawa, M., Aoki, T., Dotani, T., Takizawa, M., Tanaka, Y.,
Yoshida, K., Miyamoto, S., Iga, S., Hayashida, K., Kitamoto, S., \&
Terada, K. 1994, \pasj, 46, 375
\reference{}Inoue, H. 1991, in Frontiers of X-ray Astronomy, ed. Y. Tanaka \& K. Koyama
(Tokyo:  Universal Academy Press), 291
\reference{}Jahoda, K., Swank, J. H., Giles, A. B., Stark, M. J., Strohmayer, T.,
Zhang, W., \& Morgan, E. H. 1996, Proc. SPIE 2808, ``EUV and Gamma Ray Instumentation
for Astronomy'' VII, 59
\reference{}Jain, R., Bailyn, C. D., Orosz, J. A., Remillard R. A., \& 
McClintock, J. E. 1999, \apjl, in press
\reference{}Levine, A. M., Bradt, H., Cui, W., Jernigan, J. G., Morgan, E. H.,
Remillard, R., Shirey, R. E., \& Smith, D. A. 1996, \apjl, 469, 33
\reference{}Makishima, K., Maejima, Y., Mitsuda, K., Bradt, H. V., Remillard, R.
A., Tuohy, I. R., Hoshi, R., \& Nakagawa, M. 1986, \apj, 308, 635
\reference{}Mitsuda, K., et al. 1984, PASJ, 36, 741
\reference{}Morrison, R. \& McCammon, D. 1983, \apj, 270, 119
\reference{}Orosz, J., Bailyn, C., \& Jain, R. 1998, \iaucirc 7009
\reference{}Remillard, R. A., Orosz, J. A., McClintock, J. E., \& Bailyn, C. D. 1996,
\apj, 459, 226
\reference{}Remillard, R. A., McClintock, J. E., Sobczak, G. J., Bailyn, C. D., Orosz,
J. A., Morgan, E. H., \& Levine, A. M. 1999, \apjl, in press
\reference{}Rothschild, R. E., Blanco, P. R., Gruber, D. E., Heindl, W. A., MacDonald,
D. R., Marsden, D. C., Pelling, M. R., Wayne, L. R., \& Hink, P. L. 1998, \apj, 496, 538
\reference{}Shakura, N. I. \& Sunyaev, R. A. 1973, \aap, 24, 337
\reference{}Shimura, T. \& Takahara, F. 1995, \apj, 445, 780
\reference{}Smith, D. A. \& RXTE/ASM teams 1998, \iaucirc 7008
\reference{}Sobczak, G. J., McClintock, J. E., Remillard, R. A., Bailyn, C. D., \&
Orosz, J. A. 1999, \apj, 520, in press
\reference{}Tanaka, Y. \& Lewin, W. H. G. 1995, in X-ray Binaries, ed. W. H. G. Lewin,
J. van Paradijs, \& E. P. J. van den Heuvel (Cambridge:  Cambridge Univ. Press)
\reference{}Watson, M. G., Ricketts, M. J., \& Griffiths, R. E. 1978, \apj, 221, L69
\reference{}Wilson, C. A., Harmon, B. A., Paciesas, W. S., \& McCollough, M. L. 1998,
\iaucirc 7010
\end{references}
\end{document}